\documentclass[pre,aps,twocolumn,superscriptaddress,amsmath,showpacs,floatfix]{revtex4}
\usepackage{graphicx}
\begin{document} 
 \title{Anomalous Diffusion with Absorbing Boundary}
 \author{Yacov Kantor}
 \email{kantor@post.tau.ac.il}
 \affiliation{School for Physics and
 Astronomy, Raymond and Beverly Sackler Faculty of Exact Sciences,
 Tel Aviv University, Tel Aviv 69978, Israel}
 \author{Mehran Kardar}
 \affiliation{Department of Physics, Massachusetts Institute of
 Technology, Cambridge, Massachusetts 02139}
 
 \date{\today}
 
\begin{abstract}
In a very long Gaussian polymer on time scales shorter that the 
maximal relaxation time, the mean squared distance travelled
by a tagged monomer grows as $\sim t^{1/2}$. We analyze such sub-diffusive
behavior in the presence of one or two absorbing boundaries and
demonstrate the differences between this process and the sub-diffusion
described by the fractional Fokker-Planck equation. In particular, we
show that the mean absorption time of diffuser between two absorbing 
boundaries is finite. Our results restrict the form of
the effective dispersion equation that may describe such sub-diffusive
processes. 
\end{abstract}
\widetext
 
\maketitle

 \section{Introduction}

The stochastic fluctuations of a broad range of physical systems 
\cite{fbmgle} exhibit
a behavior commonly denoted as {\em anomalous diffusion}. 
The random motion is characterized by the scaling of a mean squared 
coordinate, which  (when averaged over many realizations) 
scales as $t^\alpha$ in time $t$.
For ``normal'' diffusion $\alpha=1$, while the
cases of $\alpha\ne1$ are referred to as ``anomalous,'' 
with $\alpha<1$, corresponding to {\em sub-diffusion},
while $\alpha>1$ describes {\em super-diffusion}. 
The physical origin of anomalous behavior is usually the coupling of
the particle (or some other coordinate) to many other degrees of freedom,
such that its dynamics is the superposition of numerous other modes
with widely distributed time scales.
In principle, there is no reason to expect any ``universality'' in
such anomalous processes, and any two situations described by the
same exponent $\alpha$ may have very different characteristics. 
Nevertheless, certain general considerations have motivated 
approaches that encompass a 
large variety of cases: for a review see Ref.~\cite{mkreview}.

Many problems related to the behavior of random walkers can be formulated
in terms of {\em first passage} of a walker, or its absorption 
at a boundary\cite{feller,redner,weiss}.
The presence of the absorbing boundary may help to discriminate 
between different types of anomalous random walkers. 
Indeed, in the following subsections we shall demonstrate how the
study of absorption can be used to gain better understanding of the
complexity of anomalous behavior.  The behavior of a normal diffuser 
confined by absorbing boundaries is well understood; 
in particular, for large times $t$ the survival probability $S(t)$ of
such a diffuser decays exponentially. Consequently, the probability 
density function (PDF) of the diffuser to be absorbed at a particular time
$Q(t)=-dS(t)/dt$ also exhibits an exponential decay, leading to a {\em finite} 
mean absorption time.  
The corresponding result in the case of sub-diffusion is less clear.
Only recently it was established  \cite{fpinf}
(while building on previously known expressions \cite{rd,mkreview})
that for one-dimensional (1D) sub-diffusion between two
absorbing boundaries {\em which is described by a particular
(fractional) diffusion equation} \cite{mkreview},  the PDF of absorption
$Q(t)$ decays as a small power of $t$, leading to an {\em infinite}
mean absorption time. 
The fractional diffusion equation used in this analysis \cite{fpinf}
applies to continuous time random walks, which at each step have
a waiting time distribution with a long tail.

A relatively simple and practically important case of sub-diffusion
is the motion of a tagged monomer in a long polymer, whose
anomalous dynamics was deduced and (numerically) observed by Kremer et 
al.~\cite{KBG}. A polymer consisting of a large number $N$ of monomers
has processes happening on multiple length scales, ranging
from the microscopic distance, such as separation between adjacent 
monomers, $a$, to the size of the polymer.
(A measure of the latter is the {\em radius of gyration} $R_g$.  
In a good solvent $R_g\approx aN^\nu$ \cite{deGennes_book},
with the exponent $\nu\approx 0.59$ in space dimension $d=3$.
The ``$\approx$'' sign indicates omission of a dimensionless prefactor
of order unity. In the absence of inter-monomer repulsion $\nu=1/2$ 
for any $d$.)
To these length scales are associated times $\tau_{\rm micro}\approx a^2/D_o$,
below which a selected monomer ``does not feel'' its surroundings,
and $\tau_N\approx R_g^2/D_{\rm CM} \approx a^2N^{1+2\nu}/D_o$ for
how long it takes 
the polymer to diffuse its own $R_g$. Here $D_o$ denotes the diffusion 
constant of a single  monomer, while  the diffusion constant of the 
entire polymer, or its center of mass (CM), is $D_o/N$. (In this discussion we disregard
hydrodynamic interactions.) Very short and very long times correspond 
to normal diffusion with different diffusion constants.
It has been shown in Refs.\cite{KBG} that for intermediate times
$\tau_{\rm micro}<t<\tau_N$, the polymer undergoes anomalous diffusion
with mean squared distance $\approx a^{2-2\alpha}(D_ot)^\alpha$, where
$\alpha=2\nu/(1+2\nu)$. Note, that sub-diffusion occurs even in the
case of the ideal polymer with $\nu=1/2$. 

In this work we analyze the sub-diffusive motion of a tagged monomer which is
part of an ideal (Gaussian) 1D polymer. While the entire polymer performs diffusive
(Monte Carlo) dynamics, we record the position of a single monomer
at the mid-point of the chain. 
In the absence of absorption, this is a simple, analytically solvable problem, 
and the exact PDF of the monomer position is easily obtained. 
We could not extend these solutions in  the presence of absorbing
boundaries, and instead resorted to numerical studies.  
Indeed, with a single absorbing boundary, the PDF of the monomer position 
{\em cannot} be found using  the standard method of images, which is 
the standard approach  
for normal diffusion and even some cases of sub-diffusion.
We show that when such a particle is placed between two 
absorbing boundaries, it has a {\em finite} mean absorption time,
which scales (as expected) with the distance between the absorbing
boundaries.
 Thus, the tagged monomer presents a simple example of sub-diffusion
 whose survival probability differs drastically from that obtained by
 application of fractional diffusion.
  
To provide the basis of comparison with anomalous diffusion,
in Sec.~\ref{sec:normdiff} we briefly review the behavior of a normal 
diffuser in the presence of absorbing boundaries. 
Our model of the tagged monomer, and the numerical procedure used,
are presented in Sec.~\ref{sec:model}. We also present some numerical 
results confirming the expected sub-diffusive motion of a single monomer. 
In Secs.~\ref{sec:single} and \ref{twoabs} we study the behavior of 
the tagged monomer in the presence of a single,
and a pair of absorbing walls, respectively. We thereby demonstrate
the similarities and distinctions between our anomalous diffuser and a normal 
random walk. Notably, we stress the differences between our case and
 the solution to the fractional diffusion equation. In the final
Sec.~\ref{sec:disc}, we discuss the possible applicability of our results 
to the the translocation of a polymer through a membrane pore, which was in
fact one of the motivations for this study.
 
\section{Normal diffusion with absorbing boundaries}\label{sec:normdiff}
 
 The simplest model of a Brownian particle is a random
walk (RW) on a discrete lattice, in which both the position of particle $R$ and 
time (number of steps) $t$ are integers.
Exact expressions for the PDF $p(R,t)$, and many other properties,
are readily available \cite{hughes}.
A continuum version is the Langevin equation for the motion (diffusion) 
of a single particle in a solvent (in the high friction limit), moving under the
influence of thermal noise \cite{kubo}:
\begin{equation}\label{langsingle}
\zeta {\partial R\over \partial t}=\eta(t).
\end{equation} 
Here $\zeta$ is the friction coefficient, and the thermal noise satisfies
$\langle \eta(t) \rangle =0$ (no bias) and
$\langle \eta(t) \eta(t')\rangle =2\zeta k_BT\delta(t-t')$. 
(The angular brackets, $\langle\rangle$, indicate averages over 
different realizations of the thermal noise.) 
Starting at $R=0$ at $t=0$, the PDF of particle position at a later time is
\begin{equation}\label{eq:gauss}
P(R,t)={1\over \sqrt{2\pi}\sigma}\exp\left(-\frac{R^2}{2\sigma^2}\right),
\end{equation}
where $\sigma^2=2D_ot$ depends on the diffusion constant $D_o=k_BT/\zeta$. 
From the Langevin equation, one can also directly construct the 
Fokker-Planck (diffusion) equation \cite{kubo,risken}  for the PDF, as
\begin{equation}\label{eq:diffusion}
{\partial P(R,t)\over\partial t}=D_o{\partial^2P(R,t)\over\partial R^2} .
\end{equation}
(Throughout this paper we consider one-dimensional motion; 
the generalization to higher dimensions is straightforward.)

Consider a diffusing particle starting at the origin 
and reaching position $R$ at time $t$ {\em without ever touching an
absorbing boundary at  $R_{\rm a}>0$}. The solution of
both continuous and  discrete versions of this problem have been 
described in detail by Chandrasekhar \cite{chandra}.
It can be shown that  the PDF for the random walker must vanish
on the absorbing boundary. Since the diffusion equation is linear, 
this boundary condition can be satisfied by superposing the PDF
of a free particle (Eq.~(\ref{eq:gauss})), and one starting at 
a {\em reflected image} as $\tilde{P}(R,t)=P(R,t)-P(R-2R_{\rm a},t)$.
The survival probability
$S(t)=\int_{-\infty}^{R_{\rm a}}\tilde{P}(R,t)$ reduced by 
absorption \cite{redner}, and for large times the absorption PDF decays as
$Q(t)\propto t^{-3/2}$. 
This PDF has diverging mean, since the particle can drift infinitely far in the
direction opposite the wall. 
It should be noted that the image method is specifically suited to
random walkers performing {\em independent unit steps};
it fails for the sub-diffusive walkers considered in this paper,
and also for long-range hops of super-diffusive motion \cite{zoia}.

If the 1D diffusing particle is confined by two absorbing boundaries, one can still
use superposition by the method of images to create a solution. However, in order to
satisfy both boundary conditions, an infinite set of images is necessary. 
A more convenient answer is obtained by expanding
the solution in terms of the eigenfunctions of the diffusion equation (\ref{eq:diffusion}). 
For a particle enclosed by absorbing boundaries at $R_{\rm a1}=0$
and $R_{\rm a2}=L$, this gives
\begin{equation}\label{eq:twoabs}
\tilde{P}(R,t)=\sum\limits_{n=1}^\infty A_n\sin\left({n\pi R\over L}\right)
\exp\left[-\left(\frac{n\pi}{L}\right)^2D_ot\right],
\end{equation}
where $\{A_n\}$ depend on the initial conditions. Note that at long 
times $\tilde{P}(R,t)\simeq A_1 \sin(\pi R/L){\rm e}^{-(\pi/L)^2D_ot}$,
i.e. the PDF has a simple sinusoidal shape with zeroes on the boundaries.
The survival probability is $S(t)\propto {\rm e}^{-t/\tau}$, where the
characteristic decay time $\tau= L^2/\pi^2D_o$ is of the order of time
the particle needs to diffuse over the length of the interval. 
The PDF for absorption, $Q(t)$, also
decays with the same time constant. 

Anomalous diffusion can in principle have a myriad of distinct causes. 
An extensively studied case corresponds to the so-called {\em continuous 
time random walks} for which the waiting time between successive steps is 
taken from a broad distribution, with power law tails and a diverging mean.  
The interest in such processes originated in studies of diffusion in 
semi-conductors\cite{pfist}, but they eventually became a prototype 
of anomalous diffusion. 
The fractional diffusion equation (which involves an integral operator) was 
developed to describe the evolution of the PDF for such walkers 
\cite{bala,schneid}, and  explicit solutions are now available 
\cite{schneid,hilfer,metz,mkreview}. 
Unlike Eq.~(\ref{eq:gauss}) the solution to these equations is not smooth,
but has a cusp at the origin. 
Another interesting feature is the behavior of the absorption PDF for
anomalous diffusers between two absorbing boundaries:
it has been shown \cite{fpinf}, by  careful analysis of the solutions
\cite{rd,mkreview}, that for large times, $Q(t)$ decays as 
$ t^{-(1+\alpha)}$,  leading, for $\alpha<1$, to a {\em diverging} 
mean absorption time!  
 
\section{The model and numerical procedure}\label{sec:model}
 
A monomer in a polymer undergoes anomalous diffusion even in the
trivial case of a {\em phantom} chain with no interactions,
for which $\nu={1/ 2}$ in any $d$. 
For this value of $\nu$, monomer fluctuations are governed by
an exponent $\alpha={1/ 2}$, and thus exhibit sub-diffusion.
Since many properties of phantom polymers can be calculated exactly,
this presents an excellent model for the study of sub-diffusion. 
For simplicity, we shall restrict ourselves to the one-dimensional 
situation; generalization to higher dimensions is straightforward.

After coarse-graining, a sufficiently long flexible polymer can be 
represented by effective monomers connected to their nearest neighbors 
by harmonic potentials (Gaussian springs)\cite{deGennes_book,Doi}.
Thus, the Hamiltonian for a chain of $N$-monomers is
\begin{equation}\label{eq:HGauss}
H={K\over2}\sum_{n=1}^{N-1}(R_{n+1}-R_n)^2.
\end{equation}
The distribution of the distance between two adjacent (along the 
chain) monomers at a temperature $T$ is governed by the Boltzmann factor 
$\exp[-\beta K(R_{n+1}-R_n)^2/2]$, with $\beta=1/(k_BT)$ and 
$k_B$ is the Boltzmann constant. The mean squared separation between 
adjacent monomers is $a^2=k_BT/K$, while the mean squared 
radius of gyration is $R_g^2={1\over 6}N(1-1/N^2)k_BT/K\approx 
{1\over 6}Nk_BT/K$.

Theoretical treatment of the polymer described above requires solution of
$N$ coupled Langevin equations. However, the problem becomes particularly
simple if we describe the configurations using {\em Rouse modes} \cite{Doi}
 \begin{equation}\label{eq:defRouse}
 U_q\equiv {1\over N}\sum_{n=1}^N R_n\cos(q(n-{1\over 2})),
 \end{equation}
where $q=p\pi/N$, and $p=0,1,\dots N-1$. In terms of $U_q$, Langevin
equations decouple, and each Rouse mode can be viewed as an independent 
``particle" moving in a harmonic potential whose strength depends on $q$. 
The PDF of every $U_q$ is Gaussian. Conversely, the position of each
monomer can be viewed a linear combination of $U_q$s (inverse of 
Eq.~(\ref{eq:defRouse})). Since the linear combination of Gaussian
variables is a Gaussian variable, we are assured that each monomer is
described exactly by a Gaussian PDF, and the theoretical study is
reduced to evaluation of the mean and variance of that distribution
(see later).
In the presence of absorption, such treatment is not possible.

 We measured distances in dimensionless units, i.e. 
multiplied by $\sqrt{K/k_BT}$. In these units the the root mean square
separation of adjacent monomers is $a=1$, and $R_g^2\approx N/6$.
We used diffusive (Rouse) dynamics to evolve the system in time, i.e. the
monomers were moved using standard Monte Carlo (MC) moves.
An elementary MC move consists of randomly picking one monomer and attempting
to increment its position by $\delta R$ chosen uniformly from the interval
(-1,+1), in dimensionless units. The change in the Boltzmann weight factor
controls the probabilistic decision of whether the move is accepted. 
$N$ elementary move attempts are defined as one MC time unit. 
The mean squared displacement of a monomer
in a single move determines the diffusion constant  $D_o$;
with the above choice of step size we had $D_o=0.10$.

\begin{figure}
\includegraphics[height=5cm]{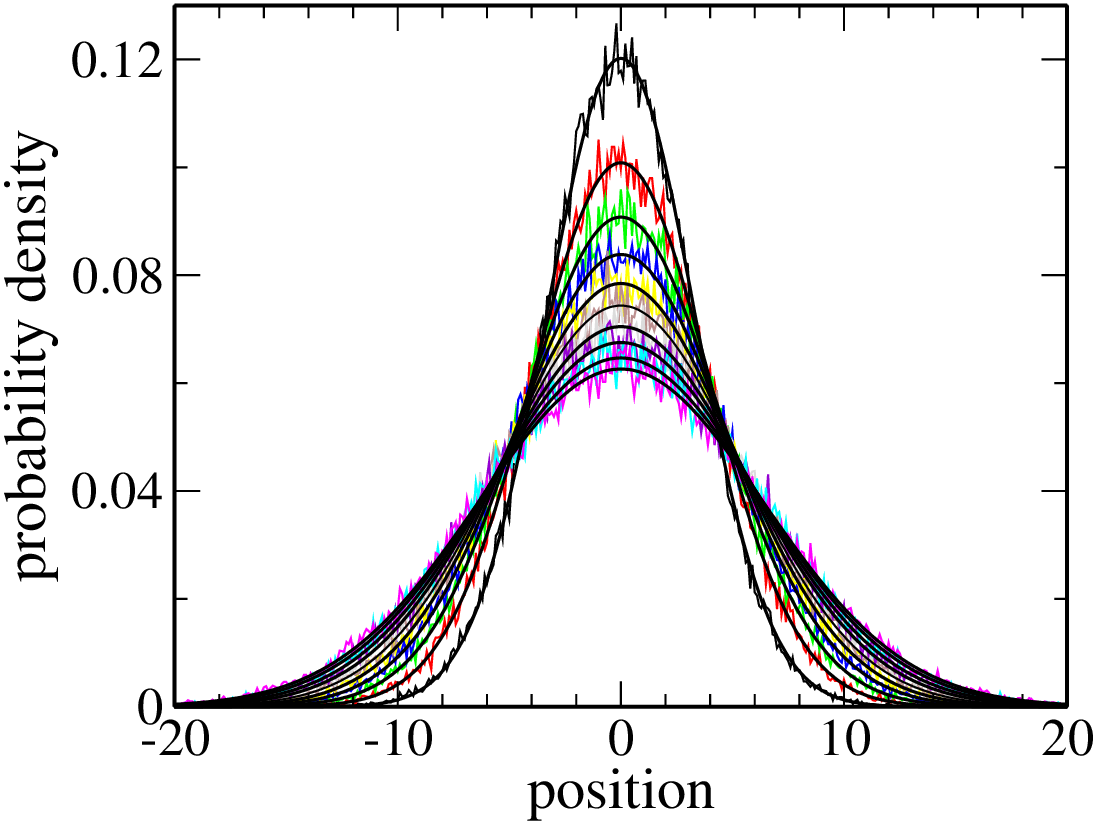}
\caption{\label{NoAbsDiff128} (Color online) PDF for the position of the central monomer in
a Gaussian chain of 129 monomers at times $t=2000, 4000,\cdots, 20000$
(from narrowest to the widest graph). The distributions were obtained
from 100,000 runs using bin size $\Delta R=0.1$.
The continuous lines represent normalized Gaussian fits to the distributions.
 }
\end{figure}

 For simulations we chose polymers of odd lengths $N=2^\ell+1$, 
with $\ell=1, 2,\cdots, 10$, i.e. $N=3, 5, 9, \cdots, 1025$. While all the
monomers moved during the simulation, we followed only the
position of the central monomer numbered $c=2^{\ell-1}+1$.
For each case, 100,000 independent simulations were performed to 
ensure reliable averages. 

As an example, Fig.~\ref{NoAbsDiff128} depicts  the PDF of the position 
of a central monomer ($c=65$) in a Gaussian polymer of $N=129$ monomers. 
At $t=0$ all monomers were located at the coordinate origin. 
As the configuration of the polymer evolved in time, the position of the 65th 
monomer was recorded. Repeating the process 100,000 times produced 
the distributions shown if Fig.~\ref{NoAbsDiff128}. Note the excellent 
(single parameter) fit of the normalized Gaussian  to the actual 
graphs. Indeed, as explained in the discussion following 
Eq.~(\ref{eq:defRouse}) the PDF of the monomer {\em must} be Gaussian
at all times. It should be noted, that this {\em shape} is the same 
as in the case  of the normal diffusion, and significantly differs from
solutions of sub-diffusive fractional diffusion equations which contain 
a cusp at the origin (see, Ref.~\cite{mkreview}). 

\begin{figure}
\includegraphics[height=5cm]{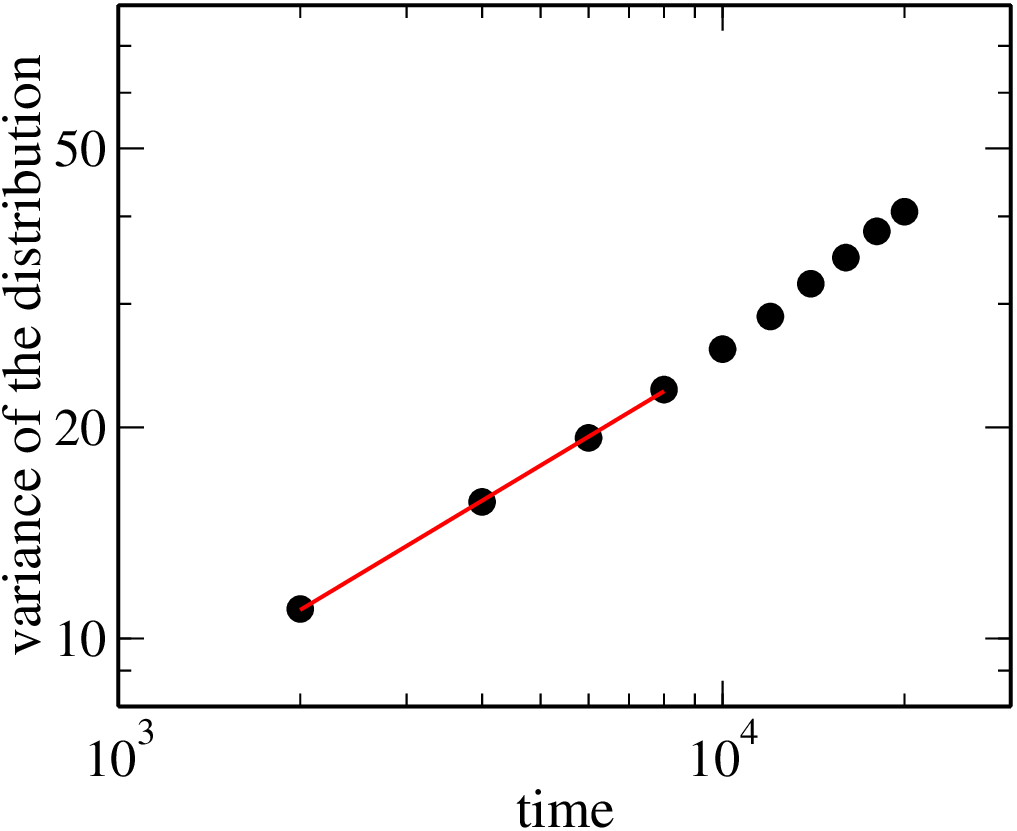}
\caption{\label{NoAbsTimeVsVar} (Color online) Logarithmic plot of the 
variance of the probability distributions
depicted at Fig.~\ref{NoAbsDiff128} as a function of time.
 }
\end{figure}

While the shapes of the graphs in Fig.~\ref{NoAbsDiff128} are not
anomalous, the time-dependence  of their variance is. 
For times shorter than the longest relaxation time 
$\tau_N=R_g^2/2D_{\rm CM}=a^2N^2/12D_o$,
which in the  above case becomes $\tau_{129}= 1.4\times 10^4$,
the variance of the distribution grows as $t^{1/2}$, while for
times longer than $\tau_N$ it is linear in $t$. This result
can be demonstrated analytically, since the variance of the
particle position can be expressed as a sum of variances of Rouse
modes. (Analogous calculation for a fluctuating line (or surface)
can be found in Ref.~\cite{kardar_book}.)
Figure~\ref{NoAbsTimeVsVar} depicts the dependence of the variance on $t$.
While all the points are within a half-decade from the crossover
point, one can clearly discern  the two types of behavior: the slope of the
straight line through the first 4 points is 0.52, very close to the
expected 1/2, and gradually increases to
the right of the graph.

\section{A single absorbing boundary}\label{sec:single}
 
Let us now introduce absorption into the  problem.
We assume that at $t=0$ all monomers  are located
at $R=0$,  and an absorbing boundary is placed at $R_{\rm a}=8$, i.e.
when the central particle reaches this point it is absorbed and
the diffusion process ends. It should be stressed that other
monomers of the polymer do not feel the absorbing boundary;
their sole function is to generate anomalous diffusion of the
tagged particle.  

We begin  with a very short polymer with $N=3$, whose
radius of gyration is significantly shorter than $R_{\rm a}$.
More importantly, its maximal relaxation time 
$\tau_{3}\approx 8$ is significantly shorter than the time 
(about $10^3$) for the CM of the polymer to diffuse the distance 
from the origin to the absorbing boundary. Therefore, at the 
time-scales at which the particle can be absorbed, the motion 
of the tagged monomer is indistinguishable from that of 
the CM of the polymer. Consequently, the problem
of absorption of the central monomer 
should be indistinguishable from that of
a normal diffuser with diffusion constant $D_o/3$.

\begin{figure}
\includegraphics[height=5cm]{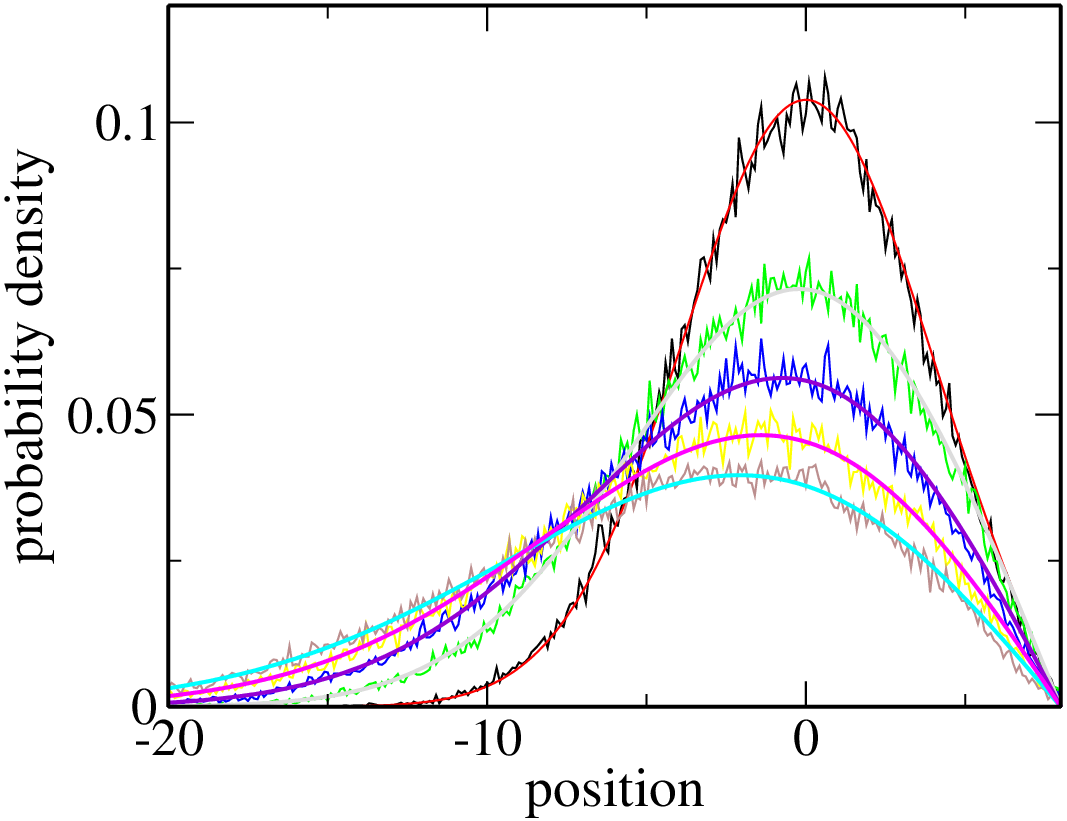}
\caption{\label{DistribWAb002} (Color online) PDF of
the position of the central particle in a polymer with $N=3$ 
monomers with an absorbing wall  at 
$R_{\rm a}=8$. The graphs correspond (narrow to broad) to $t=200, 
400,\cdots 1000$, and solid lines represent fits to the difference 
of two identical Gaussians centered at $R=0$ and 
$R=16$. The distributions are obtained from 100,000 runs,
and the bin size is $\Delta R=0.1$.
}
\end{figure}

Figure~\ref{DistribWAb002} depicts the observed PDF of the position of 
the central monomer of a polymer with $N=3$ at various times.
The area under the graphs decreases with time due to absorption.
The shapes are the same as expected for  normal
single particle diffusion: the solid lines are (single parameter)
fits to a difference between two Gaussians
centered at image points $R=0$ and $R=2R_{\rm a}$. 
The excellent fits demonstrate  that normal diffusion well describes the absorption
for such small $N$. 
 Moreover, the variances of the Gaussians fits 
increase linearly with time, with a 
prefactor corresponding to 
$\sigma^2(t)=2D_{\rm CM}t$, in which $D_{\rm CM}$ was calculated 
independently.

\begin{figure}
\includegraphics[height=5cm]{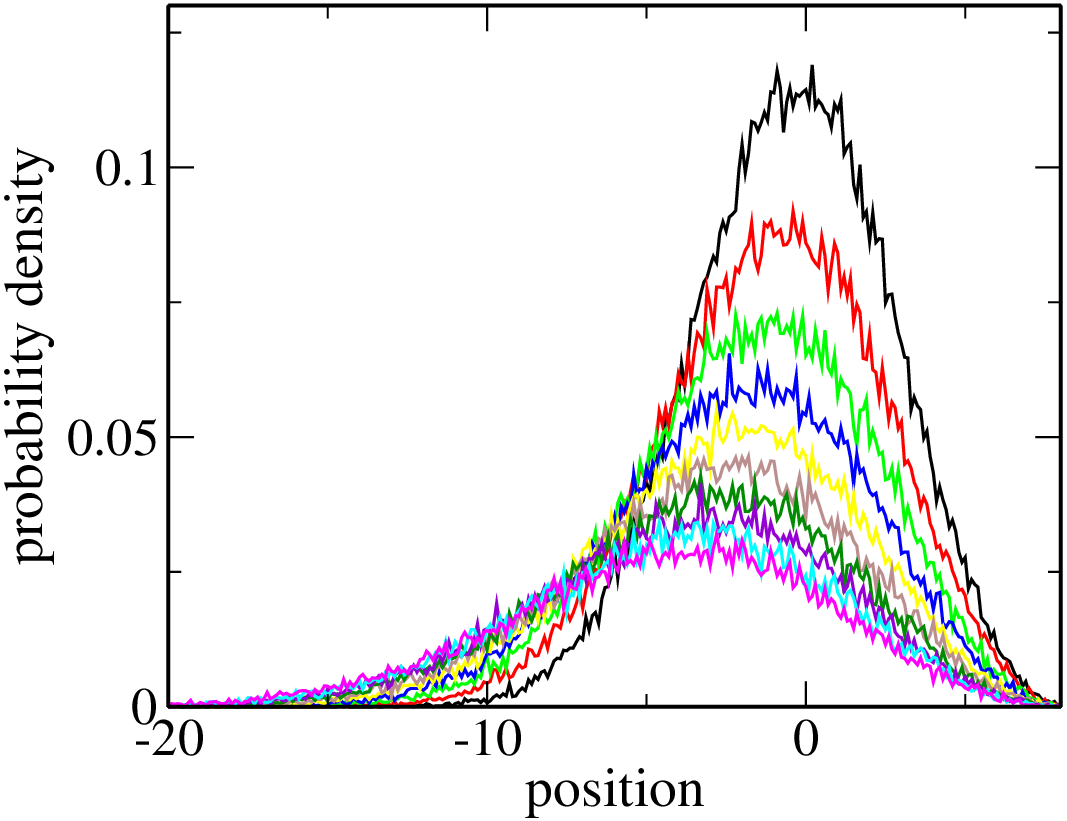}
\caption{\label{DistribWAb128a} (Color online) PDF of
the position of central particle in a polymer with
$N=129$ monomers with an absorbing wall 
at $R_{\rm a}=8$. The graphs correspond (narrow to broad) 
to times $t=2000, 4000,\cdots, 20000$. The distributions were obtained
from 100,000 runs using bin size 
$\Delta R=0.1$.
}
\end{figure}

 This behavior changes radically when $N$ becomes large. Already
for $N=129$ all resemblance to regular diffusion vanishes. The 
maximal relaxation time $\tau_{129}=1.4\times 10^4$ is of the same
order as the time required for the CM of the polymer to diffuse 
the distance to the absorbing boundary (about $4\times 10^4$), 
and $R_g$ of the polymer is of the order of the
 distance to the 
absorbing boundary. The PDF 
depicted in  Fig.~\ref{DistribWAb128a} cannot be fitted by the 
difference between two Gaussians at image points.
In particular, the PDF is not linear close to the boundary, but appears
to vanish quadratically.
Thus the qualitative behavior changes drastically on going from
regular diffusion for small $N$ to anomalous diffusion at large $N$.

\section{Two absorbing boundaries}\label{twoabs}
 
\begin{figure}
 \includegraphics[height=5cm]{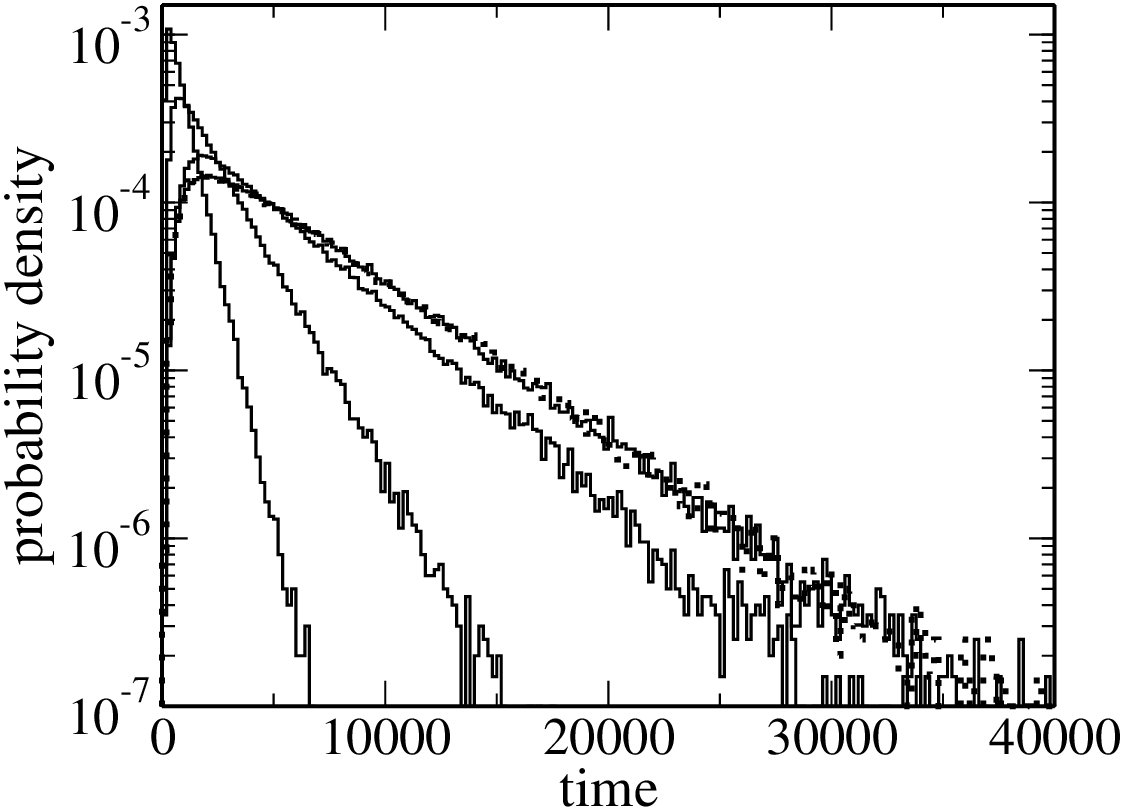}
\caption{\label{probdens} PDF of the absorption time
(in Monte Carlo time units) of the central monomer  in
 a Gaussian polymer of length $N$. The tagged monomer can be absorbed 
 at one of  two boundaries at $R_{\rm a}=\pm 8$. 
 The plots (from left to right, solid lines) correspond to $N$=3,  9, 33, 129, 
 and $N=513$ (dots), and the histogram was
 calculated from  100,000 independent runs, with bin size
 $\Delta t=200$.
 The graphs for $N=257$,  1025 (not shown), and 
 for $N=129$, are
 virtually indistinguishable.
 }
 \end{figure}

\begin{figure}
\includegraphics[height=5cm]{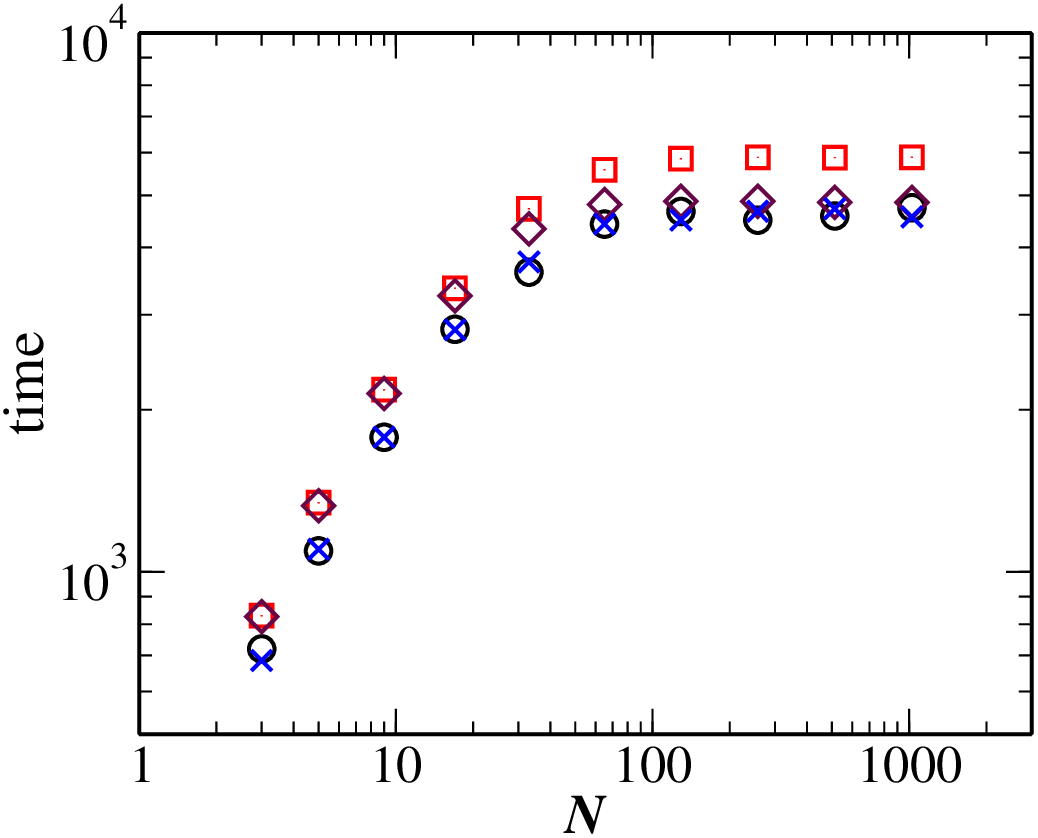}
\caption{\label{timevsL} (Color online) The mean first passage (absorption) time
(square), and the decay time-constant of the absorption PDF (circle), as a function of
polymer size $N$, when all monomers were located at
the origin at  $t=0$. Diamonds and Xs show the same quantities 
when at time $t=0$ only the central monomer is at the origin, while the
remaining monomers are in a typical equilibrium position. The data 
are obtained from 100,000 independent runs. The statistical error
bars in the mean times (approximately 0.3\%), and the estimated systematic errors 
in the decay time-constants
(less than 5\%) are much smaller than symbol sizes.}
\end{figure}

 We next consider absorption of the central monomer by two boundaries
located at $R_{\rm a}=\pm 8$, for $N$
ranging from 3 to 1025. Figure~\ref{probdens} depicts on a 
semilogarithmic scale the PDF of the absorption $Q(t) $ for several $N$.
For small polymers the curves are indistinguishable
from  that of a normal random walker
with diffusion constant $D_{\rm CM}=D_o/N$, which 
can be calculated from Eq.~(\ref{eq:twoabs}), with $\{A_n\}$ selected 
to correspond to the initial state ($P(R,0)=\delta(R)$).
After integrating $\tilde P(R,t)$ over $R$ to obtain $S(t)$, 
we get $Q(t)=-dS/dt$. As explained in
Sec.~\ref{sec:normdiff}, for large times $Q(t)$ decays
exponentially with a time constant 
$\tau\approx R_{\rm a}^2/D_{\rm CM}=NR_{\rm a}^2/D_o$. The mean absorption
time is of the same order of magnitude. The dependence of the mean time,
and of the time constant for decay, is depicted by squares and circles,
respectively in Fig.~\ref{timevsL}. 
Note that when $N$ becomes
large enough, so that the longest relaxation time of the polymer exceeds
the typical time it takes for a particle to travel the distance between
the absorbing boundaries,  $Q(t)$ becomes independent of $N$.
Indeed all the graphs for $N=129,257,513,1025$ coincide with each other.
The long time behavior remains an exponential decay, as can be seen from
the straight lines on the semi-logarithmic plot.
These curves thus depict true anomalous diffusion in the ``infinite-$N$
limit,'' and the corresponding exponential decay time constant scales as
the time it takes to cover the interval by sub-diffusion, 
 i.e. $\tau\approx R_{\rm a}^4/(a^2D_o)$.
 
\begin{figure}
\includegraphics[height=5cm]{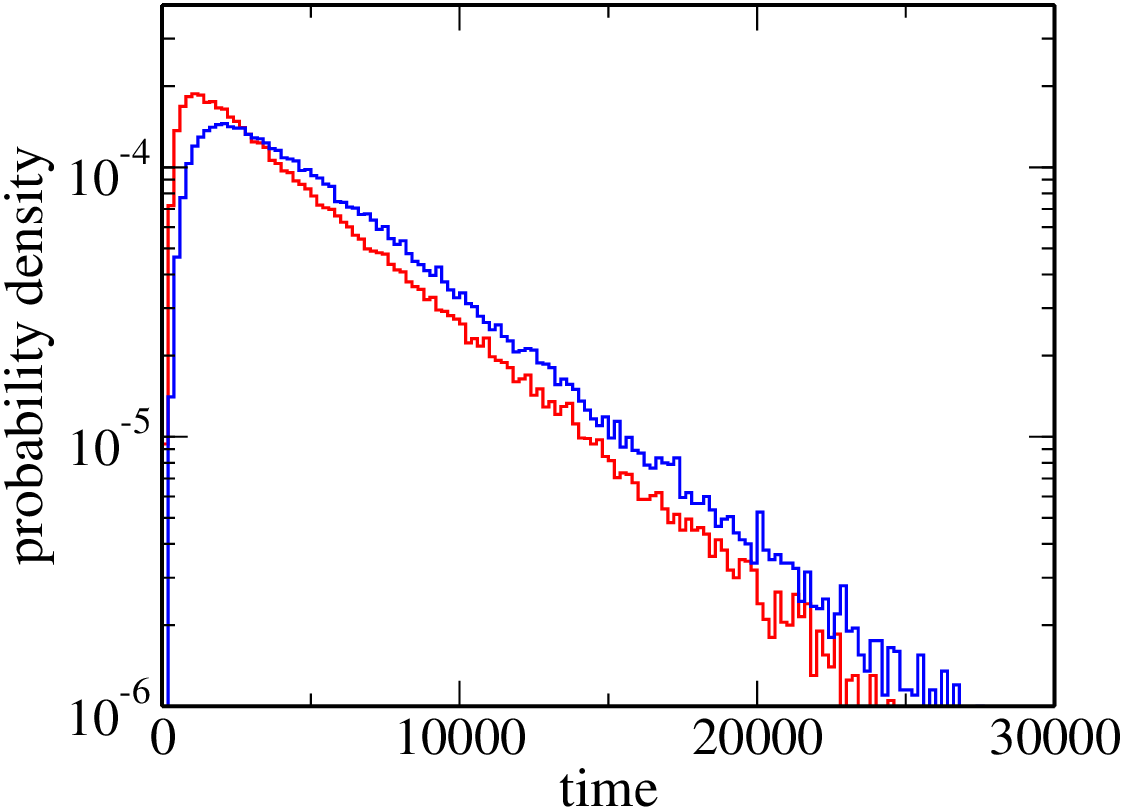}
\caption{\label{0vsEquil128} (Color online) PDF of the central monomer of
a 129-monomer Gaussian polymer absorbed at 
boundaries a distance 8 from the initial position on both sides.
The curve with the larger mean corresponds to the case when at
$t=0$ all $R_n=0$, while the other plot corresponds to starting
configuration where $R_c=0$ while the remaining monomers are 
equilibrated. This plot (histogram) was
calculated from 100,000 independent samples; bin size
$\Delta t=200$.}
 \end{figure}

So far, we reported on simulations in which at time $t=0$, the entire
polymer is located at the origin, i.e. $R_n=0$ for all $n$. 
This is a particularly convenient choice
for analytical calculations, since all Rouse modes vanish at $t=0$
and their mean values (averaged over realizations of the noise)
remain zero at all times. In any case, we know that the initial value
of each Rouse mode will be forgotten after one relaxation time of that
mode. One may consider a different case, where at $t=0$ the polymer 
assumes a randomly selected  equilibrium configuration. Thus, in addition
to averaging our results over different realizations, we also need to 
average over the starting configurations. This initial condition 
appears more natural since the time $t=0$ is not special. 
In any case, we find that the differences between the two procedures 
are rather small. For small
$N$ we cannot expect much difference, because by the time the polymer
reaches the absorbing boundary it is equilibrated in any case. The 
results for mean absorption time and decay time constant are
depicted by diamonds and Xs, respectively, in Fig.~\ref{timevsL}.
We see that the new procedure gives slightly shorter mean absorption
times and essentially the same decay times. Figure~\ref{0vsEquil128}
depicts the PDF of absorption times for $N=128$. It seems that at short
times the random starting point diffuser moves slightly faster leading
to shorter mean times, but for large times both cases are
characterized by the same decay constant. 

\begin{figure}
 \includegraphics[height=5cm]{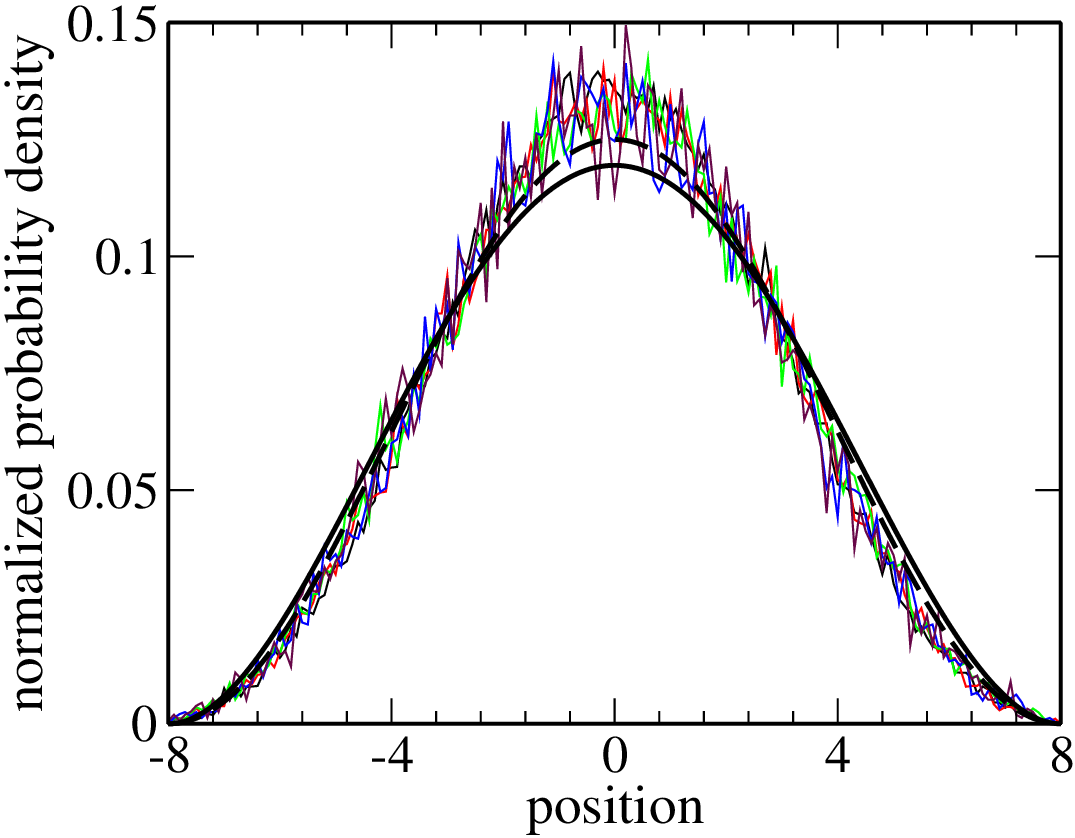}
 \caption{\label{normalized1282abs} (Color online) Probability density of 
 the position of the central monomer of a 129-monomer polymer, between absorbing
 walls located at $R_{\rm a}=\pm8$. The curves are obtained at times
 $t=2000, 4000, 6000,8000, 10000$, and are the result of 100,000 
 independent simulations collected in bins of size 
 $\Delta R=0.1$. To remove the effect of absorption all curves have
 been normalized, such that their integral is unity. The $t=2000$ curve is
 relatively smooth, while that of $t=10000$ is very noisy because
 only in a small portion of runs the tagged particle was not
 absorbed, and consequently the sample size is small. Smooth lines
 represent two suggested eigenfunctions $\Psi(xR_{\rm a})/R_{\rm a}$:
 the dashed line corresponds to $\Psi(x)=(1+\cos(\pi x))/2$ (see text), while
 the solid line represents the normalized eigenfunction given by Eq.~(A4) in 
 Ref.~\protect{\cite{zoia}}.
 }
 \end{figure}

With its exponential decay the long time behavior of a particle between
absorbing boundaries more resembles normal diffusion, although the 
time scales have to be determined using anomalous diffusion arguments. 
Nevertheless, the PDF of the
unabsorbed monomer at long times does not resemble that of 
a normal diffuser. We studied the PDF of the positions of 
surviving particles in 100,000 independent runs for a
polymer with $N=129$. Naturally, as the
time increases the probability of not being absorbed decreases.
(The decrease in probability also means that for large $t$ the
PDF was derived from samples significantly smaller than 100,000,
and consequently the statistical accuracy of the results decreased.)
To enable a convenient comparison between the PDFs at various times
we normalized them to 1. In the results depicted in 
Fig.~\ref{normalized1282abs}, the PDFs of the particle position
were recorded at different times, all of the order of
mean absorption time. Superficially these results resemble 
regular diffusion. In analogy to Eq.~(\ref{eq:twoabs}), it appears
as if at very long times only a slowest ``eigenmode" survives and
the PDF decays as $\Psi(R/R_{\rm a}){\rm e}^{-t/\tau}$, with
an eigenvalue related to $\tau$. The eigenfunction $\Psi$, 
depicted in Fig.~\ref{normalized1282abs}, appears to be
universal, although specific to our form of sub-diffusion,
while $\tau$ scales as $R_{\rm a}^4$ and is independent of $N$.

In the case of a regular diffusion (in the
scaled variable $x=R/R_{\rm a}$) we have
$\Psi(x)\propto \cos(\pi x/2)$, i.e. the function vanishes at 
the boundaries ($x=\pm1$) with a finite slope. By contrast, the results
depicted in Fig.~\ref{normalized1282abs} suggest a vanishing slope at
the boundary. In fact, an attempt to fit the function by a  few terms 
of the Fourier series $b_0+b_1\cos(\pi x)+\dots$
gives $b_0\approx b_1$ while the coefficients of higher Fourier
components are by an order of magnitude smaller. 
In Ref.~\cite{zoia} the fractional Laplacian operator was examined
in a bounded domain. Their particular 
implementation of boundary conditions enabled calculation
of the eigenfunction $\Psi(x)$ for various values of the fractional order. In
particular,  the explicit expression for the case 
corresponding to sub-diffusion with $\alpha=1/2$ in our notation
is given in Eq.~(A4) of Ref.~\cite{zoia} with
$\alpha=4$ in their notation.
While this (normalized) function, depicted by a smooth solid line in 
Fig.~\ref{normalized1282abs}, qualitatively resembles the numerical curves,
it does {\em not} provide a quantitative fit. 
This makes the fractional Laplacian operator a somewhat unlikely candidate
for describing the long-time behavior of our diffuser.

\section{Discussion}\label{sec:disc}

In this work we concentrated on an extremely simple, and yet
non-trivial model of sub-diffusion. The Gaussian nature
allows analytic calculation of some properties, such as the probability
distribution of  freely moving particles; but the absorption properties
were studied numerically. 
We believe that similar results should apply to self-avoiding polymers, 
although separation into independent  Rouse modes is no longer possible, 
and probably not much can be done beyond simple scaling arguments. 
Within our simple model, we find that the process of absorption is quite 
different from other simplified sub-diffusive processes in the literature.  
 
 Anomalous diffusion of a monomer has several features 
resembling the translocation of a long polymer through a 
narrow pore in a membrane. This process has been extensively
studied experimentally during the last decade~\cite{kasian,akeson,mel}. 
In the theoretical description
of translocation, a single variable $s$ representing the
monomer number at the pore~\cite{muthu,Lubensky,park,chern} 
indicates how much of the polymer has passed to the other side. 
If the  translocation process is very slow, the mean force 
acting on the monomer in the hole can be determined from a 
simple calculation of entropy, and the translocation problem 
is reduced to the escape of a `particle' (the translocation 
coordinate) over a potential barrier. Such theories produce
qualitative understanding of experimental results 
\cite{meller_review}.  However, if the process is {\em not} slow 
enough, compared to the relaxation times of Rouse
modes, then its dynamics is more complicated. Successive
steps of the reaction coordinate are then correlated in a manner
closely resembling the correlations between steps
of a tagged monomer in a polymer. 
In Ref.~\cite{ckk}, it was numerically verified that 
$s$ indeed undergoes anomalous diffusion in the 1D ``space
of monomer numbers.'' It was further argued that the
relaxation of the polymer constrains the translocation process
and consequently determines the translocation time. Such
behavior closely relates the translocation process to the
anomalous diffusion of a single monomer.

In the last few years significant progress has been made in the
theoretical modelling of the translocation process. On one hand,
short time behavior has been modelled in great detail~\cite{mathe},
and on the other hand scaling consideration of the long time behavior 
have been extended to include hydrodynamic interactions~\cite{storm,tian}.
Recently Grosberg et al.~\cite{GrosNech} developed an intuitive scaling
picture of polymer translocation under the influence of a force.
(See also Ref.~\cite{sakaue}.)
A variety of scaling regimes with force applied to the end-point or at the 
pore have been investigated numerically in some detail \cite{luo}.
Some recent studies~\cite{panja,dubb} suggest that the translocation
process maybe even slower than dictated by the relaxation of the Rouse 
modes. If so, this would weaken the analogy between the translocation and
the anomalous diffusion of a monomer. 
(The accuracy of these claims is questioned in further work \cite{luocom}.)

To the extent that one may draw an analogy between translocation and
anomalous diffusion in the
presence of absorbing boundaries, one may inquire whether
the mean translocation time is finite. Reference~\cite{lua} argues that 
translocation may be described by a fractional 
diffusion equation, and consequently require an infinite mean time, as found
in the solutions of such equation~\cite{fpinf}. A similar point is
made in Ref.~\cite{dubb}, where a detailed study of the PDF
of translocation times is fitted to a slowly decaying function
for large times. However, direct (experimental and numerical) 
measurements appear to indicate well defined average translocation times.
Our results offer a model where absorption times of an anomalous 
diffuser are finite. Clearly more detailed studies of translocation
are needed to resolve this question.

 In this work we studied in detail the sub-diffusion of a tagged
monomer in a Gaussian polymer. In the absence of absorption all
properties can be derived analytically. However, upon inclusion of absorbing
walls, we had to resort to numerical simulations. While we can
characterize all properties of the numerical results, we are
still missing an equation that can describe the evolution of the 
PDF of the position of the tagged particle. In fact, the numerical
results exclude several simple forms for such an equation.
 
 \begin{acknowledgments}
This work was supported by the National Science
 Foundation Grant No. DMR-04-26677 (M.K.)
 \end{acknowledgments}

 \end{document}